\documentclass[reprint,onecolumn,amsmath,amssymb,jap,aip]{revtex4-2}
\usepackage[version=3]{mhchem}
\usepackage{graphicx}
\usepackage{dcolumn}
\usepackage{bm}
\usepackage{epsfig}
\usepackage{mathptmx}
\usepackage{times}
\usepackage{booktabs}
\usepackage{dcolumn}
\usepackage{units}
\usepackage{upgreek}
\usepackage{tabularx}
\usepackage{todonotes}
\usepackage{threeparttable}
\usepackage[space]{grffile}
\usepackage[normalem]{ulem}

\usepackage{subfig}
\usepackage{multirow}
\usepackage{array}
\usepackage{xcolor}
\usepackage{soul}
\usepackage[]{caption}
\usepackage{physics}
\usepackage[T1]{fontenc}
\usepackage{units}
\usepackage{xr}
\usepackage{tocbibind} 
\usepackage{charter} % palatino} 
\makeatletter
\begin{document}

\title{Understanding the Growth and Properties of Sputter-Deposited Phase-Change Superlattice Films}
\author{Simone Prili}\affiliation{IBM Research -- Europe, S\"{a}umerstrasse 4, 8803 R\"{u}schlikon, Switzerland}\affiliation{Department of physics, University of Rome “Tor Vergata”, Via della Ricerca Scientifica 1, 00133 Rome, Italy}\affiliation{Institute for Microelectronics and Microsystems (CNR–IMM), Via del Fosso del Cavaliere 100, 00133 Rome, Italy}\email{prili@roma2.infn.it,ase@zurich.ibm.com,ghs@zurich.ibm.com}
\author{Valeria Bragaglia}\affiliation{IBM Research -- Europe, S\"{a}umerstrasse 4, 8803 R\"{u}schlikon, Switzerland}
\author{Vara Prasad Jonnalagadda}\affiliation{IBM Research -- Europe, S\"{a}umerstrasse 4, 8803 R\"{u}schlikon, Switzerland}
\author{Jesse Luchtenveld}\affiliation{IBM Research -- Europe, S\"{a}umerstrasse 4, 8803 R\"{u}schlikon, Switzerland}\affiliation{Zernike Institute for Advanced Materials, University of Groningen, Nijenborgh 4, 9747 AG Groningen, Netherlands}
\author{Bart J. Kooi}\affiliation{Zernike Institute for Advanced Materials, University of Groningen, Nijenborgh 4, 9747 AG Groningen, Netherlands}
\author{Fabrizio Arciprete}\affiliation{Department of physics, University of Rome “Tor Vergata”, Via della Ricerca Scientifica 1, 00133 Rome, Italy}\affiliation{Institute for Microelectronics and Microsystems (CNR–IMM), Via del Fosso del Cavaliere 100, 00133 Rome, Italy}
\author{Abu Sebastian}\affiliation{IBM Research -- Europe, S\"{a}umerstrasse 4, 8803 R\"{u}schlikon, Switzerland}
\author{Ghazi Sarwat Syed}\affiliation{IBM Research -- Europe, S\"{a}umerstrasse 4, 8803 R\"{u}schlikon, Switzerland}

\maketitle

\noindent{\textbf{Highly textured chalcogenide films have recently gained significant interest for phase-change memory applications. Several reports have highlighted that programming efficiency improves in devices featuring superlattice stacks, such as Ge$_2$Sb$_2$Te$_5$/Sb$_2$Te$_3$. However, to be technologically relevant, these films must be deposited on foundry-scale wafers using processes compatible with back end of the line (BEOL) integration and complementary metal-oxide-semiconductor (CMOS) technology, such as, for example, sputter deposition.
In this work, we present our observations on the influence of temperature, pressure, and seeding layer parameters on the sputter growth processes of superlattice films. By measuring various material properties, we construct a pseudo-phase diagram to illustrate the growth of both individual and superlattice films with different periodicities on technologically relevant substrates, namely SiO$_2$ and carbon. These results provide important insights into the structure, intermixing and electro-optical properties of superlattice films, and identify optimal growth parameters critical for the  manufacturability via sputtering of the material}.}

\begin{flushleft}
 \textbf{Keywords}: Phase Change Materials, Superlattice Films, Nanodevices
\end{flushleft}

\section*{Introduction}

\noindent In addition to data storage, phase-change memory devices (PCM) are increasingly being applied in analog in-memory computing\cite{sebastian2018tutorial_brain_insipred_computing,LeGallo2023_64coresIMC}. Conventional devices are usually based on untextured films of Ge-Sb-Te (GST) alloys, which are limited in performance by modest programming currents and device non-idealities. A unique solution does not exist and different approaches can be taken either at the device architecture or material level \cite{burr2016recent_dev_PCM,ghazi2021projected_mushrooms,koelmans2015projectedPCM}. In the latter case, a promising solution could be found in the use of textured chalcogenide superlattices (CSL) as the active material of the device. A superlattice (SL) is defined as periodic stack of  two or more materials which, in case of PCM, are usually chalcogenides like GeTe, Ge$_x$Sb$_y$Te$_z$ (GST) and Sb$_2$Te$_3$. The first investigations on GeTe/Sb$_2$Te$_3$ SL showed that the introduction of these materials into standard devices notably reduced the programming currents. Such reduction was initially explained supposing Ge atoms reorganization within the GeTe layers \cite{interfacial_PCM_Simpson}, leading to a crystalline-to-crystalline transition rather than the typical amorphous-to-crystalline transition via melt quenching. However, subsequent works have shown that GeTe reacts with Sb$_2$Te$_3$ during deposition, forming GST and  switching via  melt-quench dynamics\cite{nanoscale2015interface, wang2016intermixing_during_growth,terebenec2021_GeTe_SbTe_SL,boniardi2019evidence_of_thermal_trans}. There is now growing evidence that the improvements in the device performance can be attributed to the layered nature of the CSL; more specifically, from the presence of van der Waals (vdW)-like gaps within the film. These gaps render nanoscaled electro-thermal confinements, from reduced cross-plane electrical and thermal conductivity \cite{Pop2021GeTe_ST_SL_thermal_properties}. Naturally, it is expected that the performance benefits improve with the density of these gaps (number of vdW layers divided by film thickness).\\

\noindent These gaps are naturally formed in materials such as Sb$_2$Te$_3$ and GST. A unit cell of Sb$_2$Te$_3$ consists of a stack of three blocks, each about \unit[1]{nm} thick and composed of covalent bonded Te-Sb-Te-Sb-Te planes. Furthermore, these blocks are separated by vdW-like gaps \cite{kooi2020chalcogenides}, resulting in a layered trigonal structure with a lattice parameter \textit{c} of approximately \unit[3]{nm}\cite{wang2019sb2te3_structure} (i.e. 3 blocks). Undesirably, Sb$_2$Te$_3$ has a low crystallization temperature, resulting in poor state retention, thus limiting its employment in device applications. In contrast, GST has a moderate crystallization temperature and, like Sb$_2$Te$_3$, its stable phase features a trigonal unit cell composed of blocks separated by vdW-like gaps. Furthermore, the mass difference of Ge with respect to Sb and Te provides a more improved thermal confinement in GST based compositions\cite{kooi2020chalcogenides}. The lattice parameters of GST can vary depending on the stoichiometry, with each block consisting of alternating Te and Ge/Sb planes\cite{kooi2002structure_t_GST}. For example, similarly to Sb$_2$Te$_3$, in case of GST 124, the trigonal unit cell is composed of three blocks of 7 Ge-Sb-Te planes. Conversely, GST 225 features a unit cell consisting in one block of 9 Ge-Sb-Te planes.\\ 

As a result, the larger block size (\unit[1.8]{nm} for GST 225) suggest that the density of vdW-like gaps in a textured GST film would be constrained. This fundamentally presents the benefits of CSL comprised of both GST and Sb$_2$Te$_3$, with however an important caveat.  If a CSL is created with periods containing incomplete blocks, such as depositing only half a block of Sb$_2$Te$_3$, intermixing with the subsequent GST block may occur. This can lead to the formation of larger GST blocks or even not layered GST, which undesirably reduces the density of gaps and affects performance improvements \cite{Pop_GST_interfaces}. Therefore, it is imperative to grow CSL of high quality in which such intermixing effects are reduced. In this context, although much work has has been done to investigate the growth of GeTe/Sb$_2$Te$_3$ CSL with various techniques, \cite{saito2020sb2te3_seed_differentT, wang2016intermixing_during_growth, kowalczyk2018_Te_desorp_SL, boschker2017_ST_onC_MBE, yoo2022ALD_CSL, behera2018sb2te3doe, yoo2023review_CSL_growth, cremer2024growth, feng2020stickier} a systematic study on the sputter growth processes of GST/Sb$_2$Te$_3$ type CSL is still lacking. This is particularly relevant, as GST-based superlattices not only show a reduced tendency to intermix compared to their GeTe-based counterparts\cite{cecchi2017GST_ST_SL}, but have also demonstrated promising results in actual devices\cite{khan2022firstGST/Sb2Te3,wu2024novelnanocompSL}.\\

\begin{figure*}[ht]
   \includegraphics[width=\textwidth]{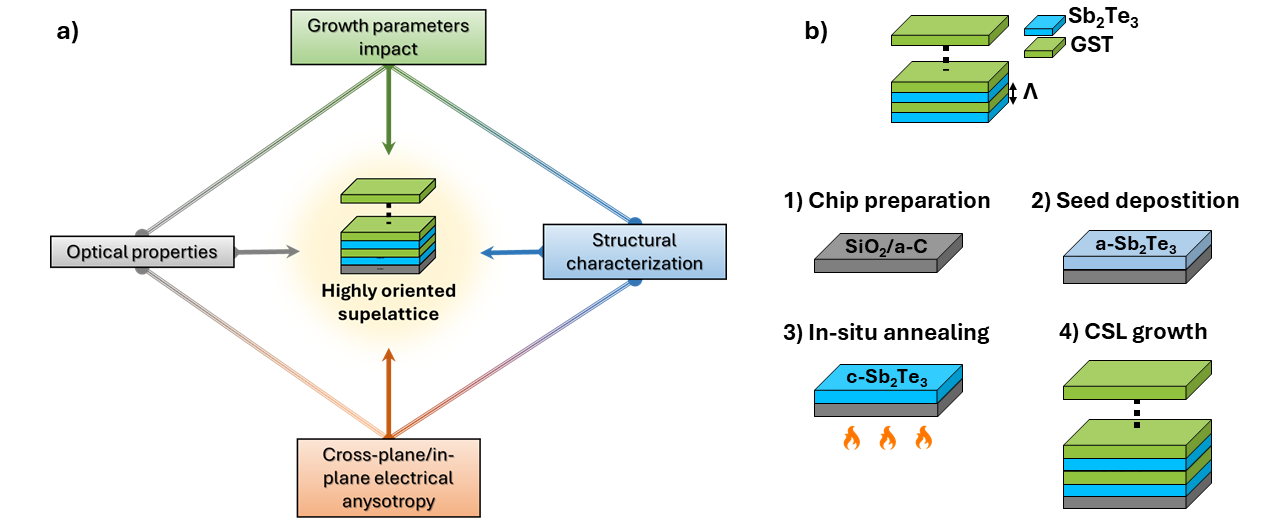}
    \centering
       \caption{\textbf{a)} An overview of the methods used to study the growth of highly oriented superlattice films; \textbf{b)} Schematics illustrating the two-step deposition scheme of the superlattice films.}
    \label{fig1}
\end{figure*}

\noindent Considering the typical use in BEOL for wafer scale deposition, here we investigate Ar$^{+}$ sputter deposition for growth of GST/Sb$_2$Te$_3$ superlattices on SiO$_2$ and C substrates (see Figure\ref{fig1}a). Specifically, we carefully separate the effects of temperature (\textit{T}) and pressure (\textit{p}), creating a \textit{p-T} diagram that serves as a framework for growing highly ordered GST and Sb$_2$Te$_3$ films, as well as CSLs with various periodicities. We utilized spectroscopic ellipsometry to monitor bandgap changes with periodicity, X-ray diffraction (XRD) for structural characterization, modified transfer length measurements (m-TLM) to analyze cross-plane and in-plane resistivities, and transmission electron microscopy (TEM) to examine the crystal structure and intermixing effects. 

\section*{Results}

\noindent We used a two-step approach for the deposition of CSL films \cite{saito2016two-step_process,saito2020sb2te3_seed_differentT,Pop2021GeTe_ST_SL_thermal_properties}. This process, as illustrated in Figure \ref{fig1}b, involves first depositing a few nanometers thick layer of Sb$_2$Te$_3$ at room temperature (seed layer). Subsequently, the sample is heated to the growth temperature (T$_{\text{growth}}$) at which the rest of the film stack is deposited. This procedure is critical for two reasons: firstly, it initiates the crystallization of the seed layer, resulting in a Te-terminated surface suitable for Sb$_2$Te$_3$ vdW epitaxy; secondly, the elevated temperature facilitates the surface diffusion of incoming atoms, thereby promoting the growth of a crystalline film. In the following sections, we develop an understanding of these growth processes. We detail the growth of Sb$_2$Te$_3$ and Ge$_2$Sb$_2$Te$_5$ in separate sections, which leads to an understanding of the optimal parameters for GST/Sb$_2$Te$_3$ CSL deposition.

\subsection*{An Understanding of S\lowercase{b}$_2$T\lowercase{e}$_3$ deposition} \label{st}
\begin{figure*}[ht]
   \includegraphics[width=\textwidth]{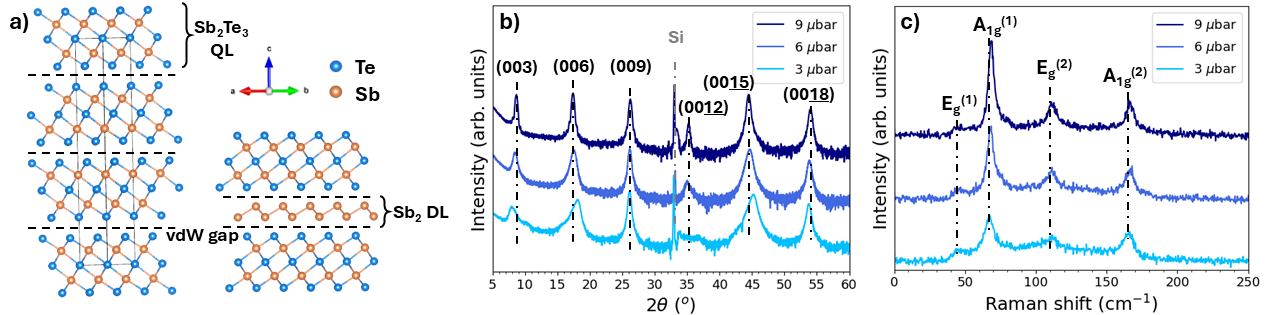}
    \centering
       \caption{\textbf{a)} Computed crystal structures of stoichiometric Sb$_2$Te$_3$ (left panel) and Sb$_2$ bilayers intercalated due to Sb enrichment (right panel); \textbf{b)} XRD patterns of Sb$_2$Te$_3$ films grown at \unit[200]{$\,^\circ\text{C}$} under different \textit{p} conditions; \textbf{c)} Raman spectra of Sb$_2$Te$_3$ films grown at \unit[200]{$\,^\circ\text{C}$} with varying \textit{p} values.}
    \label{fig2}
\end{figure*}

\noindent As illustrated in Figure \ref{fig2}a, the crystal structure of stoichiometric Sb$_2$Te$_3$ consists of vertically stacked Te-Sb-Te-Sb-Te quintuple layers (QLs) separated by vdW gaps. However, during high-temperature growth, desorption kinetics may be favored, leading to the formation of non-stoichiometric films. In the case of Sb$_2$Te$_3$, this often results in Sb-rich films that include Sb$_2$ double layers (DLs) intercalated between the Sb$_2$Te$_3$ QLs, as shown in Figure \ref{fig2}a. This deviation can affect both the properties of the alloy as well as the characteristics of the superlattices\cite{kowalczyk2018_Te_desorp_SL}.\\

\noindent In the first experiment, the effect of a seed layer on the growth of Sb$_2$Te$_3$ was evaluated by fabricating Sb$_2$Te$_3$ films at T$_{\text{growth}}$ = \unit[200]{\,$^\circ\text{C}$}, with and without a \unit[3]{nm} thick seed layer (this is detailed in Supporting Information Section \ref{SI_ST_seed}). We found that films without seed layers exhibited spurious orientations (non-(00$\ell$) diffraction peaks). Therefore, the absence of Te-terminated surfaces was noted to be detrimental to the epitaxial growth of Sb$_2$Te$_3$, indicating the fundamental requirement for a seeding layer. In the second experiment, the effect of seed layer thickness (up to \unit[5]{nm}) was investigated. We observed \unit[3]{nm} to be the optimal thickness for the seed layer, as the intensity of the diffraction peaks, indicative of higher crystal quality, diminished with increasing thickness (see Supporting Information Section \ref{SI_ST_seed}).\\
 
\noindent To investigate how Te desorption is related to sputter growth parameters, we grew Sb$_2$Te$_3$ films under different pressures \textit{p} (\unit[3/6/9]{$\mu$bar}) and analyzed their structure using XRD, as shown in Figure \ref{fig2}b. We make a few observations. The first is that the introduction of a seed layer consistently results in Sb$_2$Te$_3$ films exhibiting $(00\ell)$-only diffraction peaks, associated with  proper orientation, regardless of the pressure. However, at lower pressures, these peaks broaden and shift: particularly the (003), (006), and (0015) reflections, while the (0012) peak disappears for \unit[3]{$\mu$bar} case. This trend can be well attributed to a enrichment of Sb, which accumulates as double layers (DLs) between Sb$_2$Te$_3$ quintuple layers\cite{cecchi2019interplay}. Interestingly, due to the vdW nature of Sb$_2$ double layers\cite{yimam2023Sb2_vdW}, the desired out-of-plane orientation remains preserved even if the overall material is no longer stoichiometric. This observation is further supported by the Raman spectra of the same films shown in Figure \ref{fig2}c. The film deposited at \unit[9]{$\mu$bar} shows clear peaks at \unit[68]{cm$^{-1}$}, \unit[112]{cm$^{-1}$}, and \unit[167]{cm$^{-1}$}, corresponding to the Sb$_2$Te$_3$ A$_{1g}^{(1)}$, E$_{g}^{(2)}$, and A$_{1g}^{(2)}$ modes, respectively, along with a small bump at $\approx$ \unit[46]{cm$^{-1}$}, which could be associated with E$_{g}^{(1)}$ mode \cite{bragaglia2016Raman_modes}. As the pressure decreases, we observe a significant reduction in the intensity of the A$_{1g}^{(1)}$ and E$_g^{(2)}$ modes, with the \unit[3]{$\mu$bar} film showing a red shift to \unit[67]{cm$^{-1}$} and \unit[110]{cm$^{-1}$}, respectively. Notably, the E$_{g}^{(2)}$ mode does not exhibit the same behavior. This particular phenomenon has been experimentally and theoretically shown to result from the introduction of Sb$_2$ layers between Sb$_2$Te$_3$ quintuple layers \cite{cecchi2019interplay}, thus verifying the formation of DL.\\

\begin{figure*}[ht]
   \includegraphics[width=\textwidth]{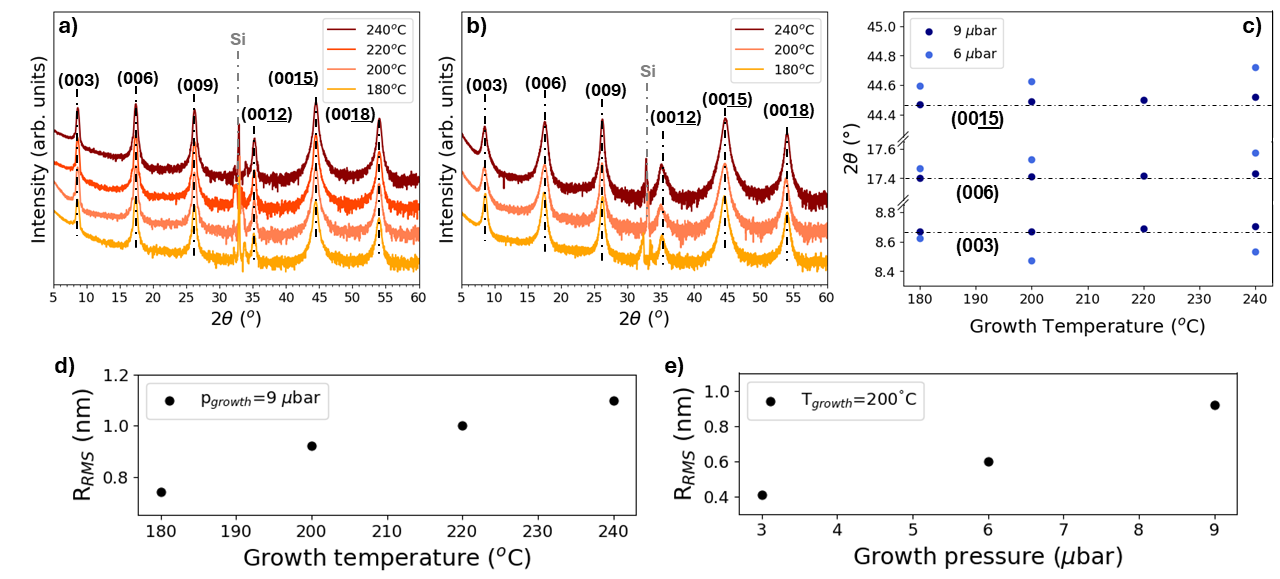}
    \centering
      \caption{XRD curves of Sb$_2$Te$_3$ films grown at \unit[9]{$\mu$bar} \textbf{(a)} and \unit[6]{$\mu$bar} \textbf{(b)} under various T$_{\text{growth}}$; \textbf{c)} $2\theta$ peak positions of $(00\ell)$ Sb$_2$Te$_3$ reflections as a function of T$_{\text{growth}}$ at \unit[9]{$\mu$bar} (dark blue) and \unit[6]{$\mu$bar} (light blue). The film roughness as a function of temperature \textbf{(d)} and pressure \textbf{(e)} .}
    \label{fig3}
\end{figure*}

\noindent To further explore Sb enrichment, we deposited Sb$_2$Te$_3$ films across a broad temperature range (\unit[180-240]{$^\circ$\text{C}}), at working pressures of \unit[9]{$\mu$bar} and \unit[6]{$\mu$bar}. The diffraction patterns of the resulting samples are shown in Figures \ref{fig3}a and \ref{fig3}b. We observed that the out-of-plane orientation in the films remains consistent regardless of the (\textit{p}, \textit{T}) conditions. However, films grown at lower pressures exhibit systematic broadening and slight shifts with increasing temperature. The (003), (006), and (0015) peaks, being the most sensitive to these variations, were fitted with pseudo-Voigt curves to determine their 2$\theta$ positions. These positions were used to monitor Sb enrichment and are plotted as a function of temperature and pressure in Figure \ref{fig2}c (dashed lines represent the 2$\theta$ positions for Sb$_2$Te$_3$ grown at \unit[9]{$\mu$bar} and \unit[200]{$\,^\circ\text{C}$}). It is evident that, within the given temperature range, pressure has an observable effect on desorption phenomena. Specifically, for films deposited at \unit[9]{$\mu$bar}, the 2$\theta$ shift from \unit[180]{$\,^\circ\text{C}$} to \unit[240]{$\,^\circ\text{C}$} is minimal and up to T $<$ \unit[220]{$\,^\circ\text{C}$}, a lattice parameter of \unit[3.0519 $\pm$ 0.0006]{nm} is observed, which is close to stoichiometric Sb$_2$Te$_3$ \cite{anderson1974refinementSbTecell}. In contrast, at lower pressures, the 2$\theta$ shift is more pronounced: films grown at \unit[6]{$\mu$bar}, \unit[180]{$\,^\circ\text{C}$} are poorer in Te compared to those grown at \unit[9]{$\mu$bar}, \unit[240]{$\,^\circ\text{C}$}. \\
\\
\noindent Given that the films must be grown on large-scale wafers, surface roughness becomes a crucial parameter. Rough films can lead to delamination issues during device fabrication and, in the case of CSL, may additionally degrade the interface quality when stacking GST and Sb$_2$Te$_3$ layers. Root mean square roughness (R$_{\text{rms}}$) obtained from AFM tomographies of Sb$_2$Te$_3$ films grown at varying \textit{p} and \textit{T} is reported in Figures \ref{fig3}d and \ref{fig3}e. It is observed that while high \textit{p} enhances Te content, it also results in rougher films. Similarly, although high \textit{T} improves film crystallinity (evidenced by more intense XRD peaks), it also leads to increased surface roughness. Consequently, we find that when working with a stoichiometric target, there is no single ideal growth condition. For instance, while it is possible to grow highly ordered, stoichiometric Sb$_2$Te$_3$ films from a stoichiometric target by minimizing Te desorption and Sb intercalation between Sb$_2$Te$_3$ QLs through optimal selection of (\textit{p}, \textit{T}) conditions, this is achieved at the expense of increased film roughness.\\

\noindent In sum, all these findings can be summarized as follows. A seed layer enhances the [001] out-of-plane orientation in Sb$_2$Te$_3$ films and reduces spurious orientations, with an optimal thickness of \unit[3]{nm}. In the typical temperature range used for growing phase change alloys, pressure is crucial for controlling Te desorption, allowing for various compositions. By fine-tuning (\textit{p} and \textit{T}), stoichiometric Sb$_2$Te$_3$ films can be produced from stoichiometric targets with controlled Sb enrichment. Higher pressure and temperature enhance film stoichiometry and crystallinity but increase surface roughness, requiring trade-offs when designing process conditions.

\subsection*{An understanding of Ge-Sb-Te deposition}
\label{gst}
\begin{figure*}[ht]
    \includegraphics[width=\textwidth]{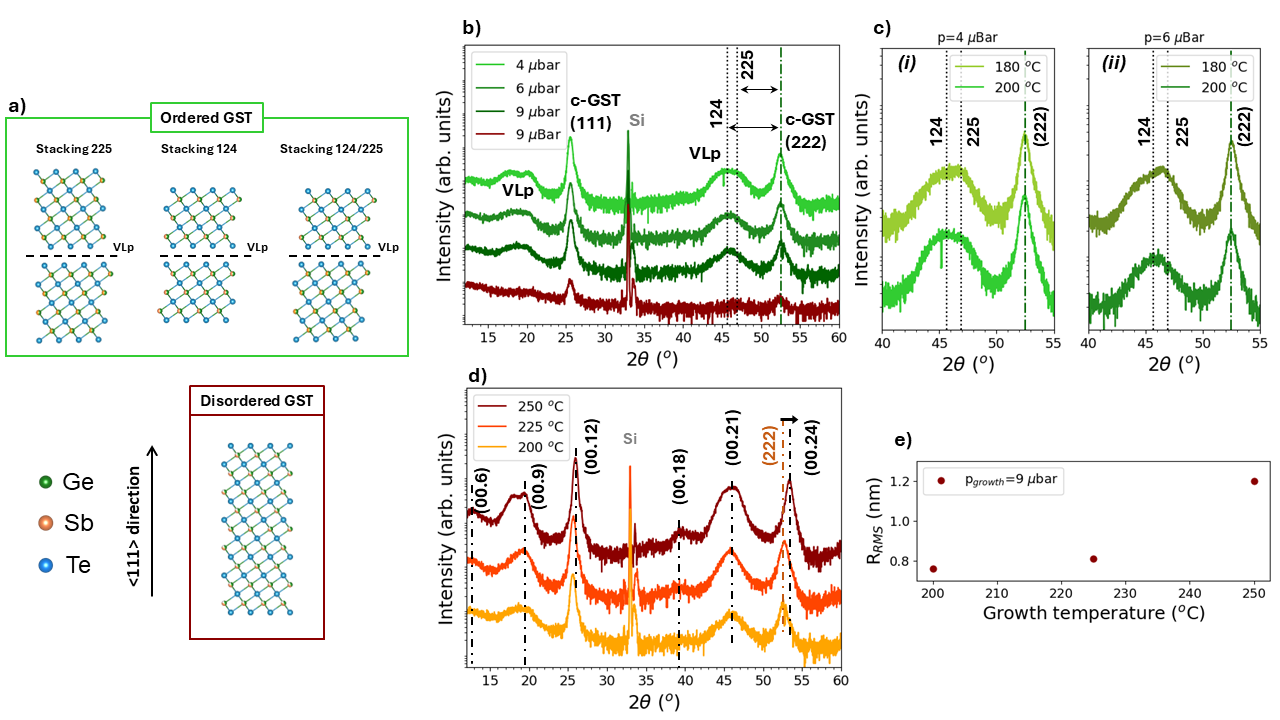}
    \centering
      \caption{\textbf{a)} Computed  structures of different cubic GST phases viewed with [111] axes in the vertical direction: ordered GST phases featuring 124, 225 and mixed 124-225 stacking (green box) and disordered cubic GST with vacancies  randomly distributed in one sublattice (red box);  \textbf{b)} XRD curves of GST films grown at \unit[200]{$\,^\circ\text{C}$} and different \textit{p} values. Red curve refer to a film grown without seed layer while green ones refer to samples featuring seed layer; \textbf{c)} Detail of GST 2$^{nd}$ diffraction order for films deposited with different \textit{T} and \textit{p} values; \textbf{d)} XRD curves of GST films grown at \unit[9]{$\mu$bar} and increasing T$_{\text{growth}}$; dashed black lines indicate t-GST reflections while (222) c-GST peak is indicated by a orange line; \textbf{e)} surface roughness measured at different values of T$_{\text{growth}}$.}
    \label{fig4}
\end{figure*}

\noindent  Besides its themodinamically stable trigonal structure, GST also features a metastable rock-salt cubic (c-) phase which is the one generally formed during electrical switching in a PCM device. It consists of alternating layers of Te and Ge/Sb/vacancies stacked along the [111] direction \cite{matsunaga2004structure_c_GST, dasilva2008insights}. Depending on the distribution of vacancies, two possible phases can be obtained: a disordered phase, where vacancies are randomly distributed in the cation sublattice, and an ordered phase, where vacancies accumulate in layers  (representing depleted Sb/Ge planes), known as vacancy layers (VL)\cite{zhang2012role_vacancies_MIT, bragaglia2016metal}. Crucially, the formation of these vacancy layers divides the c-GST into distinct blocks, with their size depending on the number of Te and Ge/Sb planes between two vacancy layers. Figure \ref{fig4}a illustrates the computed structures of disordered c-GST and ordered c-GST, showing different stacking configurations. Unlike Sb$_2$Te$_3$, the formation of VL in c-GST is understood to precede the creation of the vdW gaps of t-GST\cite{siegrist2011disorder_MIT,zhang2012role_vacancies_MIT,bragaglia2016metal}.\\

\noindent In the first experiment, we investigate the role of the Sb$_2$Te$_3$ seed layer in the growth of GST, as well as the effect of deposition pressure. XRD curves of GST films deposited at T$_{\text{growth}}$ of \unit[200]{$\,^\circ\text{C}$} under different pressures, with (green curves) and without (red curves) the seed layer, are shown in Figure \ref{fig4}b. Seeding the substrate surface with Sb$_2$Te$_3$ clearly results in higher quality crystalline films, as indicated by the overall increased intensity of the diffraction peaks. The peaks at $\simeq$ \unit[25.6]{$\,^\circ$} and \unit[52.7]{$\,^\circ$}, detected in all cases, identify the as-grown films as cubic GST with complete [111] out-of-plane orientation. Furthermore, while GST deposited without a seed layer grows in the disordered phase, films grown with a seed layer exhibit distinct peaks at approximately \unit[19]{$\,^\circ$} and \unit[45.5]{$\,^\circ$}, indicating the formation of VL\cite{bragaglia2016metal}.\\

\noindent Using this data, we further estimate (this is detailed in supporting information section \ref{SI_GST_fit}) the predominant GST stacking-type by measuring the separation, $\Delta q$ (where $q = 4\pi \sin(\theta)/\lambda$), between the (222) GST reflection and the vacancy layers peaks (VLp). This is computed as $2\pi / \Delta q$ \cite{bragaglia2018jap_designing, bragaglia2016metal,cecchi2024thick}. We note that although the films were grown using a stoichiometric GST 225 target, the vacancy layer peak (VLp) for films grown at \unit[200]{$\,^\circ\text{C}$} under \unit[9]{$\mu$bar} and \unit[6]{$\mu$bar} yield $\Delta q$=\unit[0.43]{\AA$^{-1}$}, which is suggestive of a predominant c-GST 124 stacking. In contrast, films grown at (\unit[4]{$\mu$bar}) exhibit significantly broader VLp peaks. This broadening is also observed for films deposited at \unit[180]{$\,^\circ\text{C}$} under\unit[4-6]{$\mu$bar} (as detailed in figure \ref{fig4}c). In these cases the VLp are fitted by two Gaussians: one with $\Delta q$=\unit[0.36-0.38]{\AA$^{-1}$}, that is compatible with c-GST 225, and another that could be linked to minor c-GST 124. Interestingly, while films grown at \unit[4]{$\mu$bar} display this mixed stacking at both \unit[180]{$\,^\circ\text{C}$} and \unit[200]{$\,^\circ\text{C}$} (see figure \ref{fig4}c(i)), under \unit[6]{$\mu$bar} (figure \ref{fig4}c(ii)) the same phenomena is observed only at \unit[180]{$\,^\circ\text{C}$}. In fact, increasing T$_{\text{growth}}$ to \unit[200-225]{$\,^\circ\text{C}$} result in films characterized by 124 stacking as can be observed in figure \ref{fig4}d. This suggests that unlike Sb$_2$Te$_3$, where \textit{p} was the most relevant parameter to influence film growth, in case of GST the growth temperature plays a crucial role as it influences the overall stacking of c-GST, favoring GST 124 at higher \textit{T}. Moreover, in case of T$_{growth}$=\unit[250]{$\,^\circ\text{C}$} former GST (111) and (222) peaks shift noticeably from their original positions (\unit[25.65]{$\,^\circ$} and \unit[52.66]{$\,^\circ$}, see Figure \ref{fig4}d) to \unit[25.93]{$\,^\circ$} and \unit[53.32]{$\,^\circ$}, respectively. This shift, along with the appearance of additional features in the diffractogram, indicates that the film grows in the trigonal rather than the cubic phase \cite{hilmi2017epitaxialGSTtrigXRD, bragaglia2014GST_on_Si_ann}. The emergence of the trigonal phase at a higher temperature aligns with its known tendency to crystallize at higher temperatures compared to the cubic phase. This observation is further corroborated by film's roughness (Figure \ref{fig4}e) as a function of T$_{\text{growth}}$. As long as the cubic phase is grown, R$_\text{{rms}}$ changes very little as function of T$_{\text{growth}}$ (R$_\text{{rms}}$=\unit[0.76]{nm}-\unit[0.81]{nm}); however, when the t-GST formation is triggered, R$_\text{{rms}}$ raises significantly to \unit[1.2]{nm} (at \unit[250]{$\,^\circ\text{C}$})).\\

\noindent Comparing these findings with the results on Sb$_2$Te$_3$, we observe that, in the case of GST, temperature rather than pressure seems to have a more significant effect on the film's structure. This is possibly due to the presence of Ge atoms during growth that have been suggested to preferentially bond with Te\cite{bragaglia2018jap_designing}. This is consistent with our results, as it suggests that the presence of Ge stabilizes Te adsorption, thereby reducing the impact of pressure on its desorption. In sum, the key observations are the following: a Sb$_2$Te$_3$ seed layer (thus, Sb$_2$Te$_3$) enables the growth of ordered c-GST, pressure does not have significant effect on the growth, but c-GST of different stacking-types, as well as t-GST can be obtained starting from a stoichiometric GST 225 target by fine-tuning T$_{\text{growth}}$. Concerning roughness, as long as the c-GST phase is obtained, the as grown films experience minor R$_\text{{rms}}$ variations as function of temperature and are smoother than Sb$_2$Te$_3$ films grown in the same conditions. On the other hand, if t-GST is grown, due to its fiber textured nature, R$_\text{{rms}}$ increases significantly.

\subsection*{An understanding of GST/Sb$_2$Te$_3$ superlattices growth}
\label{csl}

\begin{figure*}[ht]
  \includegraphics[width=\textwidth]{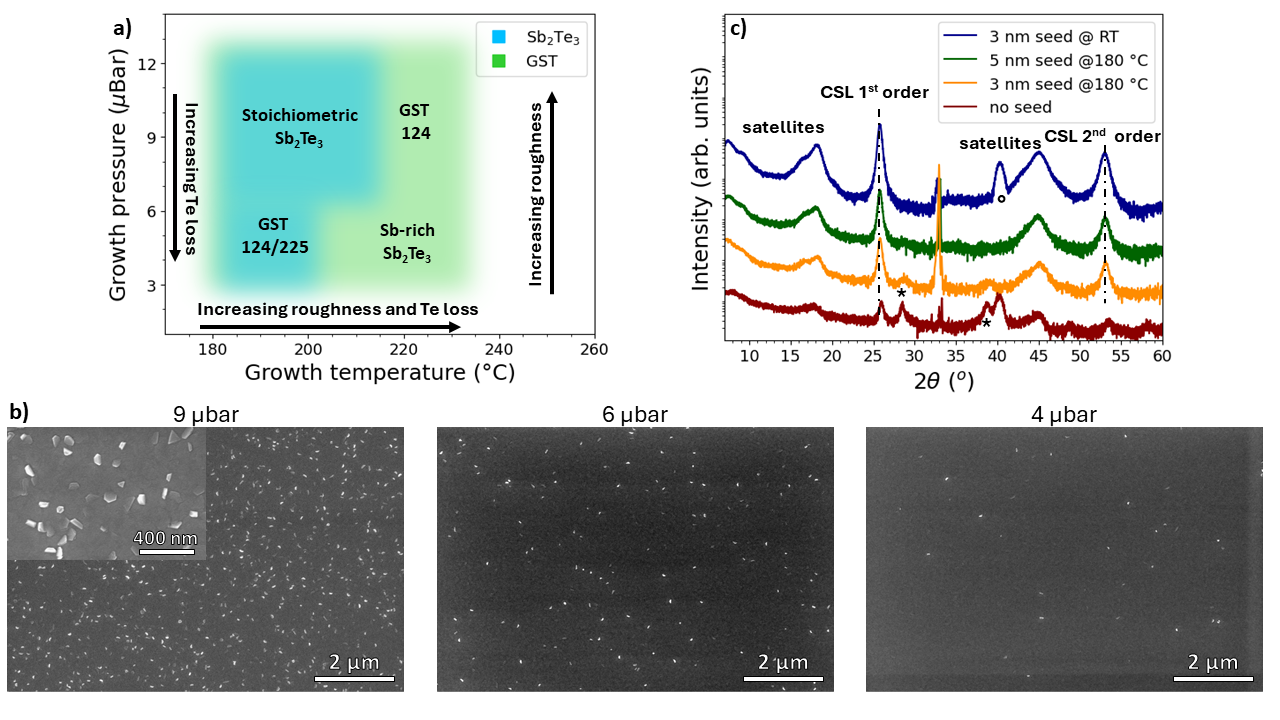}
    \caption{\textbf{a)} (\textit{p,T}) phase diagram illustrating growth conditions for GST (green) and Sb$_2$Te$_3$ (blue) films; \textbf{b)} SEM images of  CSL films grown at different p$_\text{growth}$ values. The inset illustrates flakes, indicative of delaminated Sb$_2$Te$_3$ and GST grains; \textbf{c)} XRD curves of CSL deposited without seed layer (red curve), with \unit[3]{nm} RT seed layer (topmost blue curve) and \unit[3]{nm} and \unit[5]{nm} thick high T$_\text{growth}$- deposited seed layer (second and third curve from the bottom respectively)}
    \label{fig5}
\end{figure*}

\noindent The observations gathered for the growth of Sb$_2$Te$_3$ and GST films can now be combined to draw a (\textit{p,T}) phase diagram (see Figure \ref{fig5}a), which can point to the conditions that can yield highly textured films with stoichiometric composition and low surface roughness (R$_\text{{rms}}$<\unit[1]{nm}). This phase diagram, also serves as the starting point for the growth of CSL films. For Sb$_2$Te$_3$, stoichiometric films can be grown at \unit[9]{$\mu$bar}  (or higher), with low dependence on T$_{\text{growth}}$. However, above \unit[220]{$\,^\circ\text{C}$}, the surface roughness exceeds \unit[1]{nm}, so this temperature is considered the upper bound for CSL deposition at this pressure. At \unit[3]{$\mu$bar}, despite the low roughness, notable Te desorption occurs, making it unsuitable. At \unit[6]{$\mu$bar}, despite moderate Te desorption, good quality films can be obtained in within the low T$_{\text{growth}}$ range. For GST, the pressure in the investigated range has a marginal effect on film quality. In terms of T$_{\text{growth}}$, except at \unit[250]{$\,^\circ\text{C}$}, the films maintain low roughness under the same \textit{(p,T)} conditions as Sb$_2$Te$_3$. This indicates that the GST growth window favorably overlaps with the region where high-quality Sb$_2$Te$_3$ is obtained, allowing for more streamlined parameter optimization for CSL deposition.\\

\noindent We start off with investigating the impact of pressure (up to \unit[12]{$\mu$bar}) on the CSL growth. Using XRD, we find (the analysis of diffraction curves Supporting Information section \ref{SI_press_effect}) that textured CSL films can be grown  under all \textit{p} conditions. However, CSL films grown at lower p$_\text{growth}$ values (eg. \unit[6]{$\mu$bar}) exhibits broader peaks. We are able to attribute this is to Te desorption in Sb$_2$Te$_3$ (see Sb$_2$Te$_3$ section and ref \cite{kowalczyk2018_Te_desorp_SL}). Nonetheless, we note that while higher p$_\text{growth}$ improves the overall structural quality, it negatively impacts CSL surface morphology. SEM images (see Figure \ref{fig5}b) show a gradual increase in number of flakes (bright features) as p$_\text{growth}$ increases. This effect is particularly evident for CSL grown at \unit[9]{$\mu$bar}, where the inset reveals numerous hexagonal and triangular flakes indicative of the trigonal structure characteristic of Sb$_2$Te$_3$ and GST. Interestingly, this phenomenon occurs only in the case of CSL: GST and Sb$_2$Te$_3$ films grown under the same \textit{(p,T)} conditions have smoother surfaces, as noted in the SEM and AFM analysis (see Supporting Information section \ref{SI_press_effect}).
\begin{figure}[h!]
    \centering
    \begin{minipage}[c]{0.45\textwidth}  
        \centering
        \begin{tabular}{ccccccc}
            \toprule
            \hspace{0.5cm} & $\Lambda$ \textbf{Sb$_2$Te$_3$/GST} & \hspace{1cm} & \textbf{E$_\text{gap}$} & \hspace{1cm} &\textbf{E$_\text{n}$}& \hspace{0.5cm} \\
            \hspace{0.5cm} &\textbf{(nm/nm)}& \hspace{1cm}& \textbf{(eV)}& \hspace{1cm}& \textbf{(eV)} & \hspace{0.5cm} \\
            \midrule
            \hspace{0.5cm} &4/1.8 & \hspace{1cm}& 0.31 & \hspace{1cm}& 1.50& \hspace{0.5cm} \\
            \hspace{0.5cm} &4/3.6 & \hspace{1cm}& 0.30 &\hspace{1cm} & 1.50& \hspace{0.5cm} \\
            \hspace{0.5cm} &6/3.6 & \hspace{1cm}& 0.32 &\hspace{1cm} & 1.56& \hspace{0.5cm} \\
            \hspace{0.5cm} &9/5.4 &\hspace{1cm} & 0.34 &\hspace{1cm} & 1.54 & \hspace{0.5cm}\\
            \bottomrule
        \end{tabular}
        \captionof{table}{Energy gap E$_\text{{gap}}$ and transition energy E$_\text{{n}}$ extracted from ellipsometry measurements of superlattices of varying periods $\Lambda$}
        \label{table_gap}
    \end{minipage}%
    \hfill
    \begin{minipage}[c]{0.55\textwidth}  
        \centering
        \includegraphics[width=0.8\textwidth]{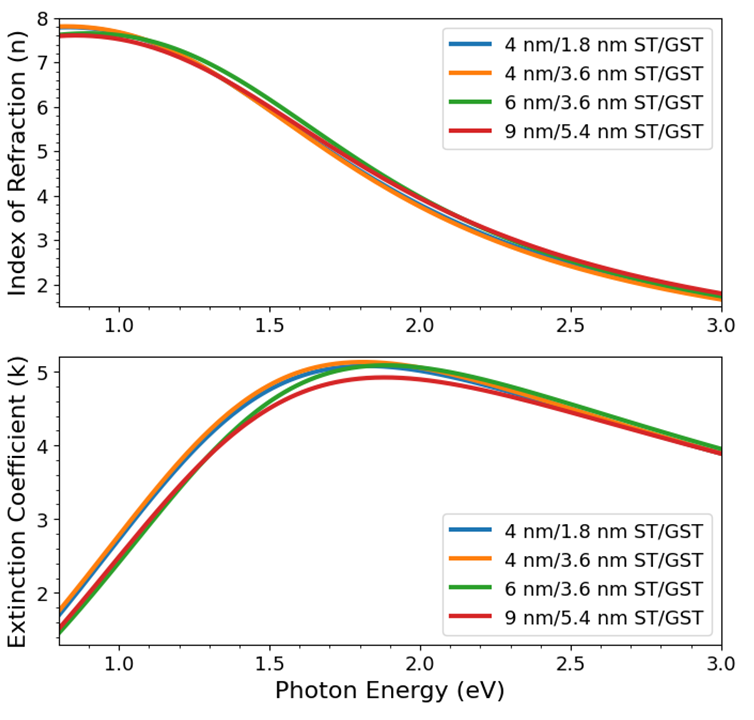}  
        \caption{Index of refraction \textit{n} and extinction coefficient \textit{k} extracted from spectroscopic ellipsometry measurements of superlattices of different periods $\Lambda$}
        \label{image_n_k}
    \end{minipage}
\end{figure}
\noindent Furthermore, the influence of the seed layer was investigated in the growth of CSL films. Figure \ref{fig5}c shows the XRD curves of superlattices (\unit[3]{nm} Sb$_2$Te$_3$/ \unit[1.8]{nm} GST) deposited at \unit[180]{$\,^\circ\text{C}$} on: bare SiO$_2$ (red), SiO$_2$ covered by \unit[3]{nm} and \unit[5]{nm} thick seed layer deposited at \unit[180]{$\,^\circ\text{C}$} (orange and green), as well as on SiO$_2$ coated by \unit[3]{nm} thick seed deposited at RT. Depositing the CSL directly on the bare substrate results in a film with a XRD curve (red) characterized by a number low intensity peaks, most of which are not related to the a superlattice structure. Peaks marked by "*" at $\simeq$ \unit[28.7]{$\,^\circ$} and \unit[39]{$\,^\circ$} originate from non (00.$\ell$) t-GST planes \cite{bertelli2022structural} while those at $\simeq$ 25.8 and 53.3 from (00.$\ell$) t-GST \cite{hilmi2017epitaxialGSTtrigXRD}  that grows without preferential orientation. The faint peaks at $\simeq$ \unit[17.5]{$\,^\circ$}, \unit[44.8]{$\,^\circ$} belong to Sb$_2$Te$_3$ (006) and (0015) planes respectively (see fig. \ref{fig2}) while the one marked by a circle is a W cap layer. In conclusion, a CSL is not obtained. On the other hand, introducing the optimized \unit[3]{nm} seed layer at RT, as observed for the films deposited at different \textit{p}, the resulting film shows a diffraction curve (blue) in which the main CSL peaks as well as its satellites are clearly observed. Undesirably, depositing an amorphous seed layer that is later crystallized through annealing can induce stresses at the substrate-film interface. This, combined with the layered nature of Sb$_2$Te$_3$, can results in film delamination during the fabrication process \cite{cohenlow, boschker2017_ST_onC_MBE}. On the other hand, since the layered nature of these materials is what we are trying to exploit, the only factor on which we can work to mitigate delamination is the RT-deposited seed.
Interestingly, depositing a \unit[3]{nm} thick Sb$_2$Te$_3$ seed layer at high temperature yields a film (XRD curve shown in Figure \ref{fig5}c) that exhibits the typical CSL features. Despite this, such peaks and satellites are less intense than those observed in case CSL deposited with the standard seed, indicating an overall lower material quality. Additionally, two minor peaks related to non (00.$\ell$) GST reflections are detected, suggesting the occurrence of intermixing phenomena that lead to formation of a fraction of non-oriented GST. These non-(00.$\ell$) features vanish if the high-T seed is deposited \unit[2]{nm} thicker, as reported in figure \ref{fig5}c, implying that with a thicker seed a better control over the layer growth is achieved. It is important to note that, even in this case, the overall intensity and definition of XRD peaks are lower compared to CSL deposited on the standard seed. This suggests that the structural and interface quality are likely worse than for CSL deposited on the RT seed. However, a layered structure with properly oriented vdW gaps is still obtained, meaning that if the RT-seed route cannot be followed due to delamination, sacrificing some structural quality could be a viable option to successfully carry out device fabrication.\\

\noindent Another interesting aspect to investigate is the relationship between optical properties and CSL periodicity. In this regard, we realized a series of CSL featuring different periodicities ($\Lambda$) on SiO$_2$ coated with RT-deposited seed layer. The as grown films were characterized via spectroscopic ellipsometry while the relevant parameters were extracted fitting the raw data with a generated model. Further details can be found in the methods section \ref{methods}.
Interestingly, regardless of the period, we measured E$_\text{{gap}}$ values falling within a rather narrow energy range (\unit[0.3-0.34]{eV}), which align with previous reports on GeTe/Sb$_2$Te$_3$ superlattices \cite{caretta2016EgapCSL, boschker2018Egap}. These values, reported in table \ref{table_gap} , are reasonable considering the bandgap energies of pure GST ($\simeq$ \unit[0.4]{eV}) and Sb$_2$Te$_3$ ($\simeq$ \unit[0.17]{eV})\cite{nvemec2009GSTellipso,park2009ST_GST_ellips}, possibly suggesting that the E$_{gap}$ of CSLs could depend on the relative amounts of GST and Sb$_2$Te$_3$ in the films, and thus within each period. Within the limits of our model, we do not observe a correlation between E$_\text{{gap}}$ and CSL period. Notably, the lowest E$_\text{{gap}}$ is obtained for $\Lambda$ = \unit[4]{nm} Sb$_2$Te$_3$/\unit[3.6]{nm} GST, which is the GST-richest superlattice we measured. A similar behavior is observed for the refractive index (\textit{n}) and the extinction coefficient (\textit{k}), as shown in Figure \ref{image_n_k}. We observed only minor variations between CSLs with different periodicities, with no clear trend emerging as a function of $\Lambda$. This indicates that factors other than the simple proportion of GST and Sb$_2$Te$_3$ might influence the optical properties of these superlattices. %whose investigation would however require more complex models (i.e. density functional theory).
\begin{figure*}[ht]
  \includegraphics[width=\textwidth]{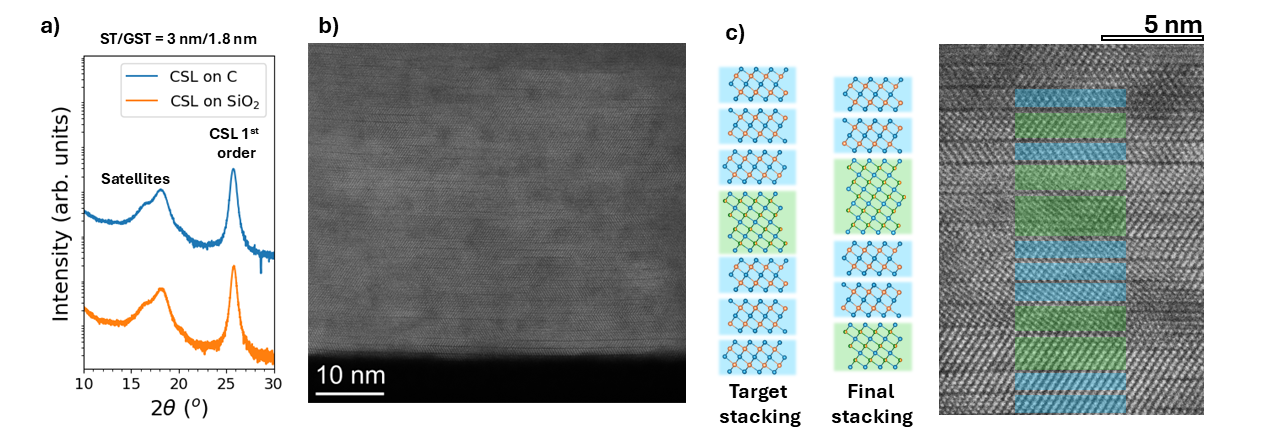}
    \caption{\textbf{a)} XRD curves of Sb$_2$Te$_3$/GST superlattices deposited on SiO$_2$ (orange) and on a C-based layer (blue) films \textbf{b)} A TEM micrograph showing the persistence of vdW gaps throughout the entire superlattice stack; \textbf{c)} Schematic representation of nominal and final CSL stacking with and without local intermixing at the interface compared with detailed micrograph %(not taken from the same area of the previous image) 
    showing Sb$_2$Te$_3$ (light blue) and GST (light green) blocks} 
    \label{fig7}
\end{figure*}
\\
\noindent We further evaluated the growth of CSL on carbon-based films, which are becoming relevant for PCM devices\cite{syed2023_C_based_memory}. To that end, CSL films 57 nm thick consisting of 12 repetitions of \unit[3]{nm} Sb$_2$Te$_3$/\unit[1.8]{nm} GST  were grown under identical \textit{(p,T)} conditions on both SiO$_2$ and the carbon-based film. The corresponding XRD curves (see Figure \ref{fig7}a) show the distinctive CSL peaks, including the main CSL diffraction order and its related satellite peaks. This indicates that CSLs can be successfully sputter-grown on carbon-based substrates. This is further evidenced on high-resolution STEM micrographs (see micrograph in Figure \ref{fig7}b), which clearly shows the presence of vdW gaps throughout the entire cross section of the film. However, the zoomed micrograph in Figure \ref{fig7}c reveals that locally the nominal Sb$_2$Te$_3$/GST periodicity is not always maintained. By counting the number of Te and Ge/Sb planes, we can identify blocks constituted by five atomic layers (Sb$_2$Te$_3$ QLs, blue) as well as blocks with 7, 9, and 11 Te/Ge/Sb planes, indicating the formation of GST124, 225, and 326 blocks. This variation is likely due to the fact that, although GST is less prone to intermixing compared to GeTe \cite{cecchi2017GST_ST_SL}, its intrinsic compositional disorder and tendency to form blocks of different sizes \cite{mio2017chemical} allow neighboring Sb$_2$Te$_3$ and GST blocks to react and form new blocks of different compositions. An example of this is schematized in Figure \ref{fig7}c, where an ideal nominal stacking sequence of \unit[3]{nm} Sb$_2$Te$_3$ (three blocks) and\unit[1.8]{nm}  GST (one 225 block) is altered by interface reactions, leading to mixed blocks. Therefore, since such intermixing could alter the nominal stacking sequence, these phenomena should be considered, especially when growing CSL of smaller thicknesses.

\begin{figure}[htp]
    \centering
    \includegraphics[width=\textwidth]{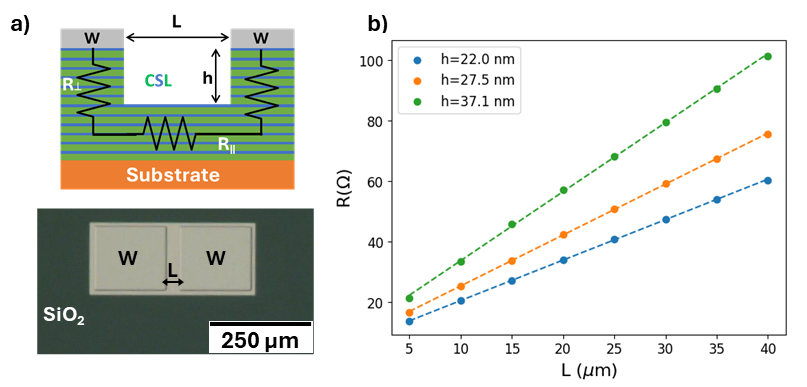}
    \caption{\textbf{a)} Cross section schematizing of the structures realized for m-TLM measurements and top view from optical microscope after fabrication; \textbf{b)} R(L) measurements (dots) and related linear fit (dashed lines) of CSL featuring different etch depth \textit{h}} 
    \label{fig8}
\end{figure}

\noindent To evaluate the electrical properties of CSL, we measured the in-plane ($\rho_{\parallel}$) and cross-plane ($\rho_{\perp}$) electrical resistivity using a modified-Transfer Length Measurements (m-TLM). Unlike standard TLM, which only estimate in-plane resistivity, m-TLM uniquely involves creating variable resistances by etching the film to different thicknesses (\textit{h}). This is depicted in Figure \ref{fig8}a. In our m-TLM structures (see Supporting Information section \ref{SI_TLM_fabr} for fabrication details), the CSL remains always capped and the contacts with electrodes are always intact. This is fundamentally different from previous studies\cite{Pop2021GeTe_ST_SL_thermal_properties}, in which the film surface is exposed to ambiance (thus, oxidation) during device fabrication.  The total resistance measured between the two contacts can be written as
$R_{\text{total}} = R_c + 2R_{\perp} + R_{\parallel}$
where $R_c$ is the contact resistance with the electrodes, $R_{\perp}$ is the cross-plane resistance from CSL, and $R_{\parallel}$ is the in-plane resistance of CSL. \\

\noindent Figure \ref{fig8}b shows an exemplar plot of $R_{\text{total}}$ as a function of the channel length (L) for a superlattice \unit[96]{nm} thick progressively etched to vary the values of \textit{h}. We firstly note that the increase in measured electrical resistance as a function of \textit{L} is perfectly linear up to \unit[40]{$\mu$m} (the maximum pad distance), indicating excellent long-range uniformity of our films. $\rho_{\parallel}$ is computed from the slopes of these $R_{\text{total}}(L)$ curves, taking the value $\rho_{\parallel} = 1.19 \pm 0.02 \times 10^{-3} \ \Omega \text{cm}$ (average of $\rho_{\parallel}$ extracted from each curve: \unit[0.99$\times 10^{-3}$]{$\Omega \text{cm}$}, \unit[1.15$\times 10^{-3}$]{$\Omega \text{cm}$} and \unit[1.44$\times 10^{-3}$]{$\Omega \text{cm}$}). This value is in line with superlattices data reported in literature, as shown in the supporting information.\\
\noindent In principle, modified TLM also allows to assess  $\rho_{\perp}$ by plotting the intercepts ($q$) of these $R_{\text{total}}(L)$ curves as a function of \textit{h}. Calculating the $\rho_{\perp}$ with this approach in the supplementary information, we obtain, as expected, a $\rho_{\perp}$ value higher than $\rho_{\parallel}$. However, the measured $\rho_{\perp}$ is also significantly higher than values previously reported in the literature. This discrepancy could result from partial oxidation of the vertical pillars in the test structures, which may have affected the measurement. Further investigations are needed to confirm this, while a discussion and a comparison to literature are provided in the supplementary information.

\section*{Conclusion}

\noindent

\noindent In summary, we have investigated the effects of growth parameters, such as pressure and temperature, on the sputter deposition of GST, Sb$_2$Te$_3$, and their superlattices (CSL). The seed layer of a specific thickness was critical for achieving oriented growth, with pressure being particularly important for Sb$_2$Te$_3$ due to its impact on Te desorption. For GST, deposition optimization enabled control over the stacking of the cubic phase, as well as the growth of t-GST, without altering the target composition, with temperature being the most influential factor. This is particularly relevant in light of renewed interest in textured GST as active material\cite{cohenlow, terebenec2024gete}. A 
\textit{p,T} diagram was developed to map optimal growth conditions for these films, resulting in highly textured and uniform growth of CSL films with varying periodicities on relevant substrates. CSL growth was also achieved using a seed layer deposited at high T, useful to minimize potential delamination. Atomic characterization revealed local intermixing at the interfaces. Characterization of the optical refractive indices showed no dependency on periodicity. The quality of interfaces could be explained by deposition defined desorption kinetics and film roughness. These results provide a directed methodology for sputter growth of CSL as well as textured Sb$_2$Te$_3$ and GST.

\small

\section*{Methods}\label{methods}

\noindent The sputter deposition of the films were carried out using FHR.Star.75.Co DC magnetron sputtering tool.  Single GST 225 and Sb$_2$Te$_3$ stoichiometric targets were used, and depositions were carried out with $\simeq$ low\unit[10$^{-7}$]{$\mu$bar} base pressure. Films were deposited on \unit[20]{mm} $\times$ \unit[20]{mm} Si(100) chips covered by $\simeq$\unit[100]{nm} thick bare, and amorphous carbon coated, thermal SiO$_2$. Prior to material deposition, each substrate was cleaned with acetone and isopropanol for \unit[10]{’} and in-situ with Ar+ inverse sputtering etching (ISE) process to remove native oxide. Deposition rates of Sb$_2$Te$_3$ and GST films were evaluated via X-ray reflectivity measurements. To maintain low deposition rate, \unit[30]{W} sputtering power was used. Depending on the subsequent processing steps, samples were capped \textit{in situ} either with SiO$_2$ or W. Out of plane diffraction curves were acquired in $\theta-2\theta$ symmetric geometry using Bruker D8 discovery diffractometer mounting Cu anode and using K$\alpha$ radiation ($\lambda$=\unit[1.5406]{\AA} ). Ellipsometry measurements were performed in a J.A. Wollam spectroscopic elllipsometer. Data collected at room temperature at three incidence angles (\unit[60]{$\,^\circ$}, \unit[65]{$\,^\circ$}, and \unit[70]{$\,^\circ$}).  Optical constants and other relevant parameters were extracted by fitting experimental data to a Tauc-Lorentz oscillator  model\cite{jellison1996parameterization, jellison1996erratum} on WVASE software package. Each sample was simulated with a layered model featuring one layer for the substrate, one for the active material and one for the capping layer and/or surface roughness. Imaging was performed using a probe- and image-corrected ThermoFisher Scientific ThemisZ S/TEM microscope operated at \unit[300]{kV} in STEM mode. A current of approximately \unit[20]{pA} was used not to destroy the phase-change material, at a convergence angle of approximately \unit[25]{mrad}. The detectors used were a high-angle annular dark-field (HAADF) detector, a four-segment annular detector (DF4) for integrated differential phase contrast (iDPC). For the images, the dwell time was set to \unit[2]{$\mu$s}  for a 4096×4096 pixel resolution. The elemental maps of 1024×1024 pixels were collected with drift correction over several hours to get sufficient counts with the low current.

%%References
\section*{Acknowledgements}\label{Acknowledgements}
\noindent This work was supported by the IBM Research AI Hardware Center. The authors would like to express their gratitude to the Binnig and Rohrer Nanotechnology Center (BRNC). S.P. would like to thank Prof. Stefano Cecchi for insightful discussions on XRD and to Prof. Fabio de Matteis for his support with ellipsometry data analysis.

\section{Supporting information}
\subsection{An understanding of the seed layer}\label{SI_ST_seed}
\begin{figure}[htp]
    \centering
    \includegraphics[width=14cm]{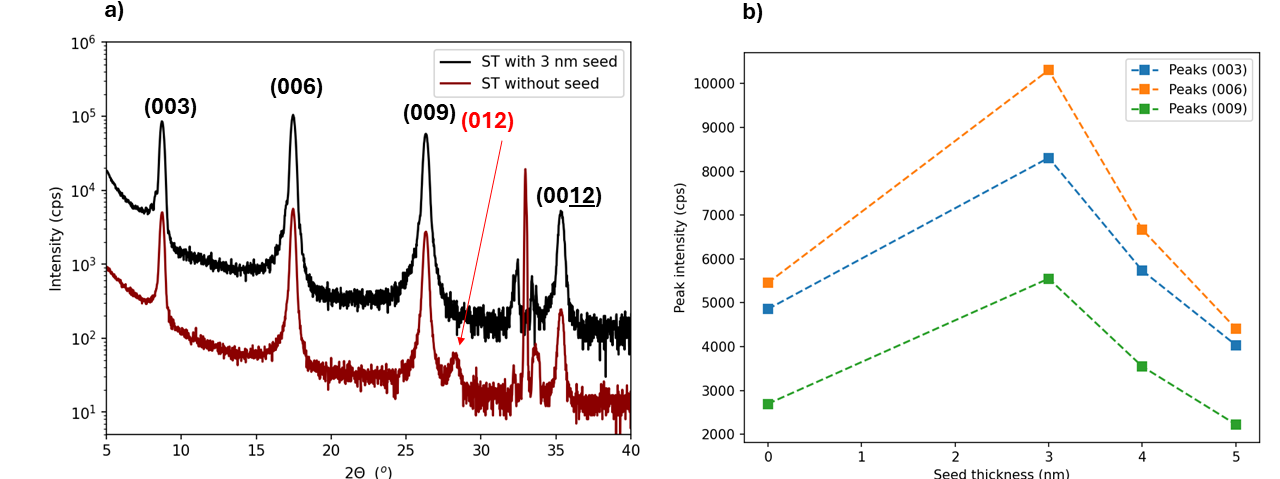}
    \caption{\textbf{a)} Sb$_2$Te$_3$ films grown in identical conditions with and without seed layer; \textbf{b)} Intensity of (00$\ell$) XRD peaks plotted as function of the thickness of the seed layer employed for the growth of the related film} 
    \label{fig_ST_seed_SI}
\end{figure}

\noindent In the Section detailing Sb$_2$Te$_3$ in the main text, we noted that using an Sb$_2$Te$_3$ seed layer enables a highly textured growth. Here we discuss this further. The effect of the seed layer was investigated by initially depositing $\simeq$ \unit[25]{nm} Sb$_2$Te$_3$ on SiO$_2$ under identical conditions, with and without a \unit[3]{nm} thick Sb$_2$Te$_3$ seed. The XRD curves of the resulting films are shown in Figure \ref{fig_ST_seed_SI}a. In both cases, crystalline films were achieved. Notably, the film deposited using the seed layer was completely out-of-plane oriented, as indicated by the presence of (00$\ell$) diffraction peaks. In contrast, when no seed layer was used, the Sb$_2$Te$_3$ still grew predominantly oriented, but a minor component of misoriented Sb$_2$Te$_3$ was also formed. This can be inferred from the appearance of the (012) peak, which is the most favored orientation in Sb$_2$Te$_3$ powders.
Figure \ref{fig_ST_seed_SI}b shows the intensity of (00$\ell$) diffraction peaks from films deposited under identical conditions but with seed layers of varying thicknesses. The data indicate that the highest number of counts (XRD peaks intensity) was achieved with a \unit[3]{nm} thick seed layer, corresponding to three quintuple layers stacked one on top of the other. This suggests that a \unit[3]{nm} seed layer provides the optimal thickness for achieving high-quality crystalline Sb$_2$Te$_3$ films.
\subsection{An understanding of the type of GST blocks}\label{SI_GST_fit}
\begin{figure}
    \centering
    \includegraphics[width=1\linewidth]{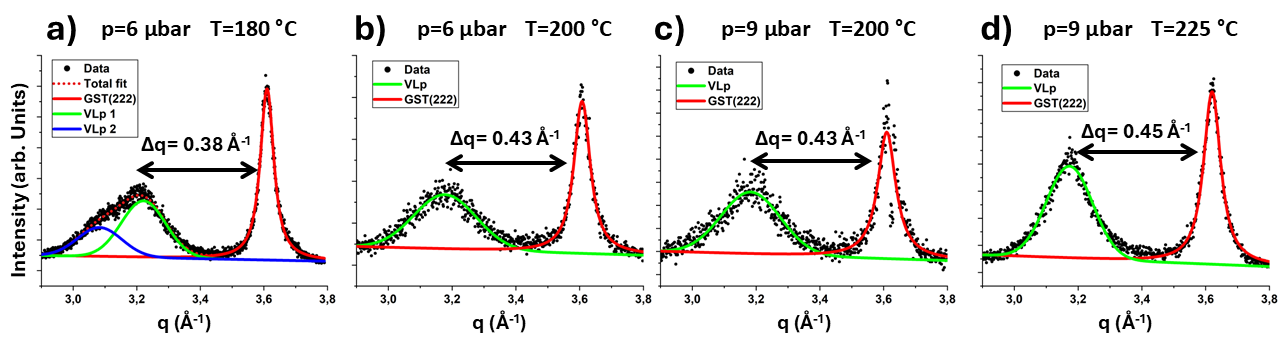}
    \caption{XRD curves, and related fits, of GST films grown in different conditions: \unit[180]{$\,^\circ\text{C}$} and \unit[6]{$\mu$bar} \textbf{(a)}; \unit[200]{$\,^\circ\text{C}$} and \unit[6]{$\mu$bar} \textbf{(b)}; \unit[200]{$\,^\circ\text{C}$} and \unit[9]{$\mu$bar}  \textbf{(c)}; \unit[225]{$\,^\circ\text{C}$} and \unit[9]{$\mu$bar} \textbf{(d)}. Red curves are pseudo-Voigts functions used to fit GST(222) reflections while green and blue curves are Gaussian for VLp modeling}
    \label{fig_XRD_GST_fits}
\end{figure}

\noindent In the Section discussing Ge-Sb-Te in the main text, we calculated the predominant stacking sequence of c-GST by fitting VLp in the diffraction curves. Here, we present the detailed description and results of these fittings. It is well established in the literature that XRD is a valuable tool for determining the predominant stacking of c-GST \cite{bragaglia2018jap_designing, bragaglia2016metal}. This involves fitting the GST(111)/GST(222) XRD peaks along with the VLp feature. Typically, the main GST features are modeled using a pseudo-Voigt function, while the VLp is fitted with one or more Gaussian curves. Given that the second-order diffraction peak of the VLp is usually more intense than the first, our analysis focuses on fitting the GST(222) peak and its associated VLp.
After fitting the XRD curve, the stacking (\textit{S}) is calculated based on the distance $\Delta q$ between the GST(222) peak and the Gaussian used to fit the VLp, according to the relation $S=2\pi / \Delta q$. The results of this analysis are shown in Figure \ref{fig_XRD_GST_fits}, illustrating the some of the cases described in the main text. For films deposited at \unit[6]{$\mu$bar}, two Gaussians are required to model the VLp of GST grown at \unit[180]{$\,^\circ\text{C}$}, while only one Gaussian is sufficient for the film deposited at \unit[200]{$\,^\circ\text{C}$}. By computing the $\Delta q$ distance between the GST and VLp peaks for the film grown at \unit[180]{$\,^\circ\text{C}$} ($\Delta q=$ \unit[0.38]{\AA$^{-1}$} and \unit[0.5]{\AA$^{-1}$}), results in blocksizes compatible with mixed 124/225-predominant 225 stacking\cite{zallo2021evolution}. Similar results are obtained with films grown under \unit[4]{$\mu$bar} at \unit[180-200]{$\,^\circ\text{C}$}. Conversely, complete 124 stacking is observed under \unit[6]{$\mu$bar} at \unit[200]{$\,^\circ\text{C}$} and under \unit[9]{$\mu$bar} at both \unit[200]{$\,^\circ\text{C}$} and \unit[225]{$\,^\circ\text{C}$}, as proved by the single-component VLp located at $\Delta q\simeq$ \unit[0.43-0.45]{\AA$^{-1}$} .

\subsection{An understanding of the effect of pressure on CSL structure and morphology}\label{SI_press_effect}

\begin{figure}[htp]
    \centering
    \includegraphics[width=8cm]{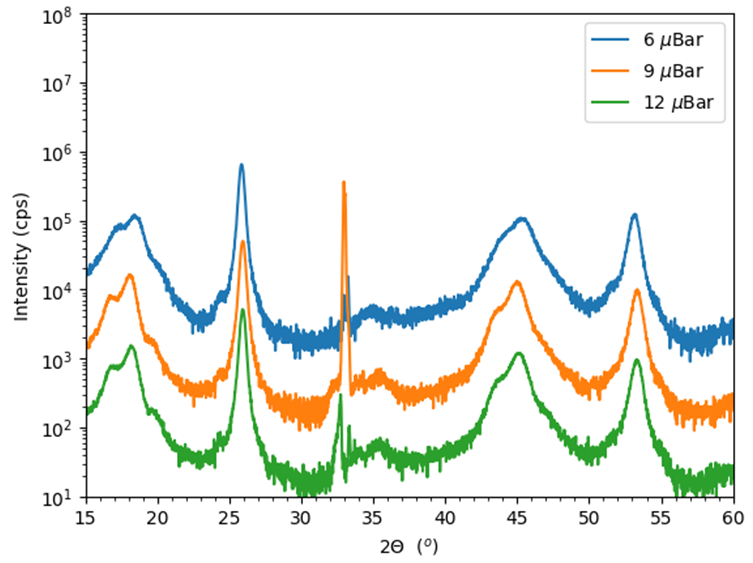}
    \caption{XRD curves of CSL grown at 200°C under different pressure conditions} 
    \label{fig_XRD_CSLvsp}
\end{figure}

\noindent In the Section detailing CSL growth in the main text, we discussed the effect of growth pressure on the superlattices. Here, we present the XRD curves along with additional AFM and SEM data.
The effect of pressure on CSL films was evaluated by growing them at \unit[200]{$\,^\circ\text{C}$} under different increasing pressures. The XRD curves reported in Figure \ref{fig_XRD_CSLvsp} show no spurious signals from segregated unoriented GST/ Sb$_2$Te$_3$, while the first and second orders of diffraction of the CSL are very clear. Qualitative information about the structural quality can be deduced from the satellite peaks observed on the left of the main CSL peaks. As the pressure decreases, these peaks change modestly from \unit[12]{$\mu$bar} to \unit[9]{$\mu$bar}, while a clear broadening is observed in the film grown at \unit[6]{$\mu$bar}. This is not surprising considering that Sb$_2$Te$_3$ experiences Te desorption at \unit[6]{$\mu$bar} while it remains stoichiometric at \unit[9]{$\mu$bar} or above. Therefore, since Te depletion causes broader CSL peaks \cite{kowalczyk2018_Te_desorp_SL}, we can attribute this broadening to Te loss in  Sb$_2$Te$_3$. 

\begin{figure}[htp]
    \centering
    \includegraphics[width=0.9\linewidth]{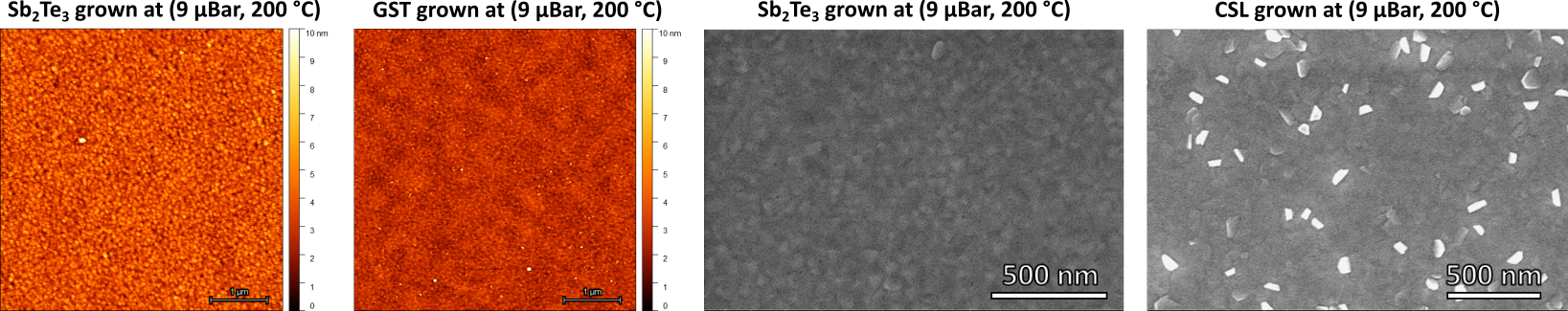}
    \caption{SEM and AFM images of Sb$_2$Te$_3$, GST and CSL grown at \unit[200]{$\,^\circ\text{C}$} and \unit[9]{$\mu$bar}} 
    \label{fig_surfaces_SI}
\end{figure}
\noindent In figure \ref{fig_surfaces_SI}, we present AFM and SEM images of the surfaces of  Sb$_2$Te$_3$, GST, and a superlattice, all grown under identical conditions.  It is clear that while the CSL shows a number of flakes, seemingly due to delamination, the GST and Sb$_2$Te$_3$ grown under the same conditions are perfectly flat, with no signs of flaking. 

\section{Fabrication of TLM devices and cross plane resistivity}\label{SI_TLM_fabr}

\begin{figure}[htp]
    \centering
    \includegraphics[width=14cm]{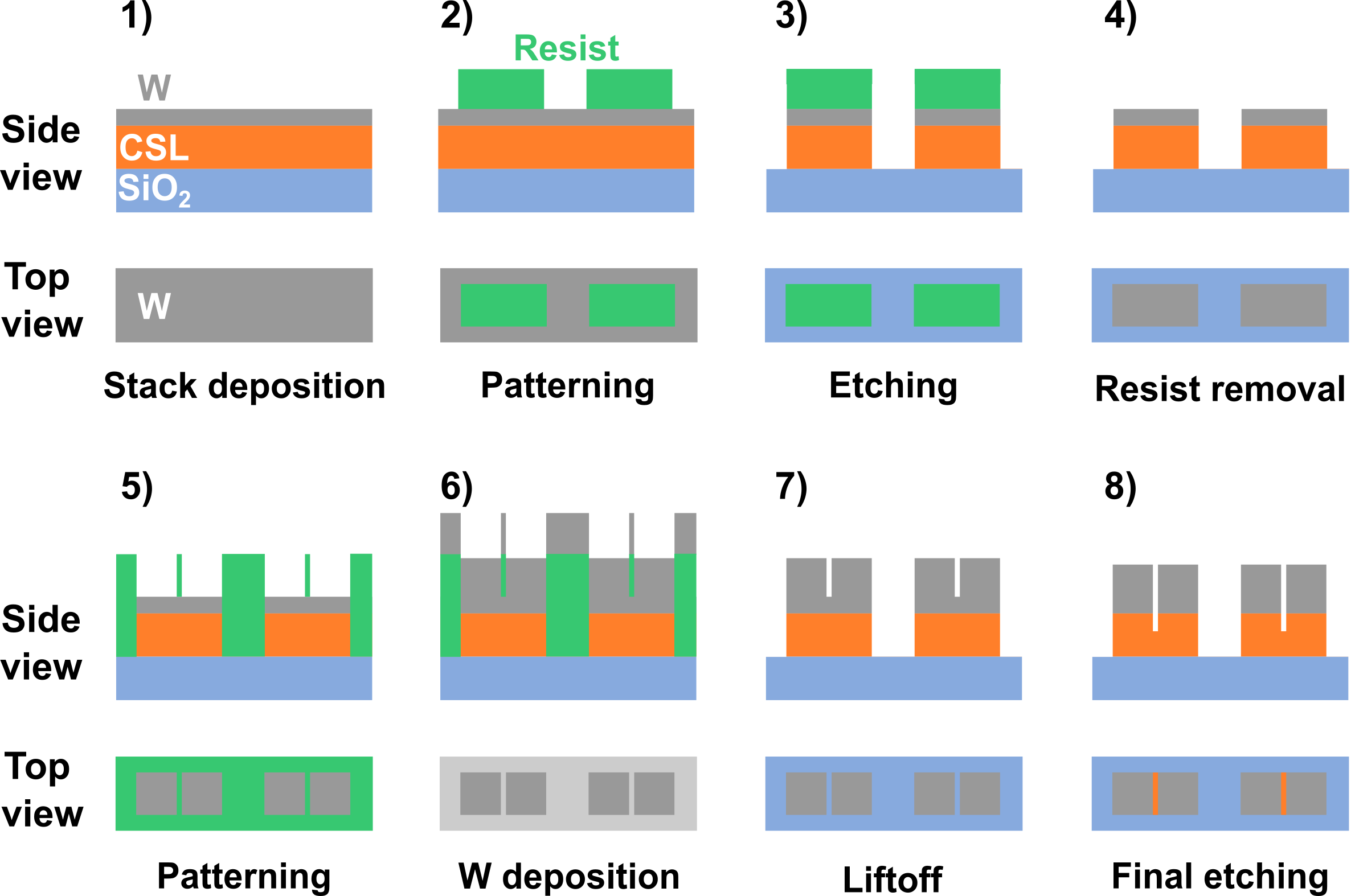}
    \caption{Schematics of the fabrication process of the test structures used for m-TLM. The figure shows for each of the 8 steps both side view (cross section) and top view of the chip. SiO$_{2}$ substrate is blue, CSL film is orange, W is gray and the negative resist is green. Pictures not in scale.} 
    \label{fig_fabrication_SI}
\end{figure}

\noindent In Section discussing electrical characterization in the main text, we presented data obtained from m-TLM to assess in-plane and cross-plane resistivity. Here, we provide a detailed description of the 8 fabrication steps schematized in Figure \ref{fig_fabrication_SI}. Initially, the substrate underwent the usual chemical cleaning procedure to prepare it for film deposition. Next, the CSL film was sputter-grown on the cleaned substrate. After allowing the film to cool down, it was capped with a \unit[10]{nm} layer of tungsten (W) deposited at room temperature (1). This W layer will be the starting point for the contacts and, since it is deposited in situ, it effectively prevents CSL surface exposure to atmosphere (thus oxidation).
Following film deposition, the chip was removed from the growth chamber for a first lithography step that employed a negative resist to define the area of the TLM test structure (2). After the resist was developed, an ion milling etching was performed to isolate the test structures (3). The resist was then removed (4).
Subsequently, another lithography step (5) with negative resist was carried out to define the pad area (\unit[100]{$\mu$m} $\times$\unit[100]{$\mu$m}). This consisted of leaving the contact area free from resist while the channel area remained covered. Subsequently, the chip was placed in a sputter tool for the deposition of\unit[100]{nm} W pads (6). Prior this deposition inverse sputter etch was conducted to remove any native W oxide that may have formed (\unit[1-2]{nm}).
The process then continued with a liftoff that removed the resists and defined the final test structures (7). A final ion milling step (8) ended the fabrication of the device, etching the remaining W in the channel as well as a portion \textit{h} of the superlattice. In this way, we fabricated 80 TLM test structures, 10 for each channel length (L=\unit[5, 10, 15, 20, 25, 30, 35, 40]{$\mu$m}).\\

\begin{figure}[h!]
    \centering
    \begin{minipage}[c]{0.5\textwidth}
        \centering
        \includegraphics[width=1\textwidth]{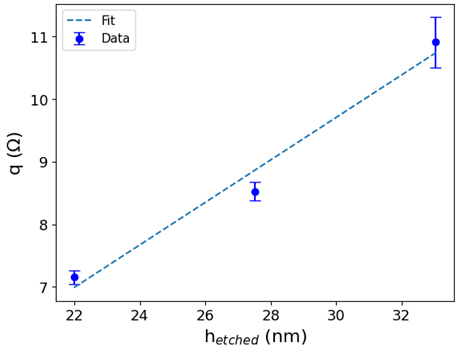}  
        \caption{Intercept \textit{q(h$_{etched}$)} as function of the etch depth \textit{h} and linear fit (dashed line) used for $\rho_{\perp}$ evaluation}
        \label{image_cp}
    \end{minipage}%
    \hfill
    \begin{minipage}[c]{0.45\textwidth}  
        \centering
        \captionof{table}{Table reporting in plane and cross plane electrical resistivity from various superlattices reported in literature grown with various techniques}
        \begin{tabular}{cccc}
            \toprule
             Material & In plane & Cross-plane & Ref. \\                 & ($\Omega$ cm) & ($\Omega$ cm) & \\
            \midrule
            GST/Sb$_{2}$Te$_{3}$ SL & 1.19 $\times$ 10$^{-3}$ & 140 & This work\\
            GeTe/Sb$_{2}$Te$_{3}$ SL & 5.8 $\times$ 10$^{-4}$ & 1.1 & \cite{Pop2021GeTe_ST_SL_thermal_properties}\\
            GeTe/Sb$_{2}$Te$_{3}$ SL & 4.8 $\times$ 10$^{-3}$ & 19.4 & \cite{yang2017ip_cp_resistivity}\\
            In$_{2}$Se$_{3}$ nanowire SL & 2 $\times$ 10$^{-3}$ & 40 & \cite{peng2008large}\\
            GST/Sb$_{2}$Te$_{3}$ SL & 9.2 $\times$ 10$^{-4}$ &  & \cite{cecchi2017GST_ST_SL}\\
            \bottomrule
            \end{tabular}        
    \label{compar_table}
    \end{minipage}
\end{figure}
\noindent For $\rho_{\perp}$, the intercepts ($q$) of the $R_{\text{total}}(L)$ curves are plotted as a function of \textit{h} and, since the intercept effectively equates to $q = 2R_c + 2\rho_{\perp} h/A$. Consequently, $\rho_{\perp}$ can be extracted from the slope of the $q(h)$ curve in Figure \ref{image_cp}, yielding $\rho_{\perp} = 169\pm36 \ \Omega \text{cm}$. This translates in a stark contrast between the cross-plane and in-plane resistivity, on the order of \unit[100000]{x}.
However, as summarized in Table \ref{compar_table}, the $\rho_{\perp}$/$\rho_{\parallel}$ ratio observed here is significantly higher than previous reports in the literature \cite{Pop2021GeTe_ST_SL_thermal_properties,  yang2017ip_cp_resistivity, cecchi2017GST_ST_SL}, primarily due to an unusually high cross-plane resistivity. While a substantial difference between in-plane and cross-plane conductivity is expected, such an elevated value could be influenced by additional factors. For instance, all the sidewalls of the vertical pillars, which contribute to the cross-plane resistivity, are directly exposed to the atmosphere, increasing the surface area susceptible to oxidation. This could compromise the accuracy of the cross-plane resistivity measurement. Therefore, to validate the reported $\rho_{\perp}$ value, further investigations are necessary to rule out the influence of such effects. Conversely, the in plane counterpart is less affected as the portion of film involved in the evaluation of $\rho_{\parallel}$ exposes much less area to the ambiance. As a result, the measured $\rho_{\parallel}$ value  is in line with previous reports (see table \ref{compar_table}).

\def\url#1{}
\bibliographystyle{naturemag}
\bibliography{PCMbiblio}

\end{document}